\documentclass[11pt]{article}
\usepackage[utf8]{inputenc}
\usepackage{titling}
\usepackage{amsmath, amssymb,amsthm}
\usepackage{graphicx}				
\usepackage{hyperref}
\usepackage{scrextend}
\usepackage{slashed}
\usepackage{braket}
\usepackage{gensymb}
\usepackage[official]{eurosym}
\usepackage[position=bottom]{floatrow}
\usepackage[letterpaper, top=1in,bottom=1in,left=1in,right=1in]{geometry}
\usepackage[backend=bibtex, style=phys, sorting=none, giveninits=true]{biblatex}
\usepackage[colorinlistoftodos]{todonotes}

\usepackage{graphicx,epstopdf}
\usepackage{amsmath}
\usepackage{float}
\usepackage{bm}
\usepackage{comment} 
\usepackage{subfigure}
\usepackage{amstext}
\usepackage{amssymb}
\usepackage{manyfoot}
\usepackage{footmisc}
\usepackage{perpage}
\usepackage{multicol}
\usepackage{paralist}
\usepackage{longtable}
\usepackage{tikz}
\usepackage{colortbl}
\usepackage{multirow}
\usepackage{ragged2e}
\usepackage{tabularx}
\usepackage{siunitx}
\usepackage[T1]{fontenc}
\usepackage{xspace}
\usepackage{comment} 

\usepackage{url}
\usepackage{hyperref}
\hypersetup{colorlinks=true, urlcolor=blue, linkcolor=blue,  citecolor=blue,  plainpages=false, breaklinks=true, bookmarksnumbered=true, bookmarksopen=true, pdfkeywords={}, pdfpagemode=UseOutlines }	
\usepackage{enumitem}
\setlist{topsep=0.5mm,parsep=0.5mm}

\newcommand{\nexp}{PIONEER}
\newcommand{\atar}{\mbox{ATAR}}
\newcommand{\calo}{\mbox{CALO}}
\newcommand{\tracker}{\mbox{TRACKER}}

\newcommand{\pie}{$\pi\to e \nu$}
\newcommand{\pimue}{$\pi\to\mu\to e$}
\newcommand{\remu}{\mbox{$R_{e/\mu}$}}

\newcommand{\pdar}{\mbox{$\pi$DAR}}

\newcommand{\pdif}{\mbox{$\pi$DIF}}
\newcommand{\mdif}{\mbox{$\mu$DIF}}

\bibliography{main.bib}

\begin{document}

\title{European Strategy for Particle Physics -- 2026 Update \linebreak PIONEER: a next generation rare pion decay experiment}

\author{PIONEER Collaboration}
\date{}
\begin{titlingpage}
\maketitle
\abstract{
PIONEER is a rapidly developing effort aimed to perform a pristine test of lepton flavour universality (LFU) and of the unitarity of the first row of the CKM matrix by significantly improving the measurements of rare decays of the charged pion. The experiment is approved at the Paul Scherrer Institute (PSI). In Phase I, PIONEER aims to measure the charged-pion branching ratio to electrons vs.\ muons $R_{e/\mu}$ to 1 part in $10^4$, improving the current experimental result $R_{e/\mu} \hspace{0.1cm}\text{(exp)} =1.2327(23)\times10^{-4}$
by a factor of 15. This precision on $R_{e/\mu}$ will match the theoretical accuracy of the SM prediction allowing 
for a test of LFU at an unprecedented level, 
probing non-SM explanations of LFU violation 
through sensitivity to quantum effects of new particles up to the PeV mass scale.

Phase II and III  will aim to improve the experimental precision of the branching ratio of pion 
beta decay, $\pi^+\to \pi^0 e^+ \nu (\gamma)$, currently at $1.036(6)\times10^{-8}$, by a factor 
of three and six, respectively. The improved measurements will be used to extract $V_{ud}$ in a theoretically pristine manner.
The ultimate precision of $V_{ud}$ is expected to reach the  0.05\,\% level, allowing for a stringent test of CKM unitarity. 

The PIONEER experiment will also improve the experimental limits
by an order of magnitude or more on a host of exotic decays that probe the effects of heavy neutrinos and dark sector physics. 

The conceptual design of PIONEER includes a 3$\pi$-sr 19 radiation length calorimeter, a segmented low-gain avalanche diode (LGAD)  stopping target, a positron tracker, and ultra-fast electronics. Compared to the previous generation of rare pion decay experiments, the 5-D (position, time, and energy) tracking capability of the LGAD-based active target allows for excellent separation of $\pi \rightarrow  e \nu$ signal from vast amount of $\pi\to\mu\to e$ background ($\pi\to \mu\nu$ followed by $\mu \rightarrow e \nu  \overline{\nu}$).
 
The PIONEER collaboration consists of participants from both the nuclear and particle physics communities
including PIENU, PEN/PiBeta, and MEG/MEGII collaborations, as well as experts in rare kaon decays, 
low-energy stopped muon experiments, the Muon $g-2$ experimental campaign, high energy collider physics, 
neutrino physics, and other areas. The collaboration is engaged in R\&D in several critical areas including i) beam studies, ii) LGAD-based active target (sensor and readout electronics), iii) calorimetry (Noble gas and crystals), iv) DAQ, and v) trigger.  A detailed simulation framework is used to estimate sensitivity and systematics.  The collaboration is still developing and welcomes new members.

This input to the 2026 update of the European Strategy for Particle Physics Strategy describes the physics motivation and the conceptual design of the PIONEER experiment, and is prepared based on the PIONEER proposal submitted to and approved with  high priority by the PSI  program advisory committee (PAC). Using intense pion beams, and state-of-the-art instrumentation and computational resources, the PIONEER experiment is aiming to begin data taking by the end of this decade

}
\end{titlingpage}

\clearpage


\section{Rare pion decays: scientific context and objectives}
\begin{figure}[h!]
\begin{center}
\begin{tabular}{cc}
\includegraphics[width=0.48\linewidth, height=160pt]{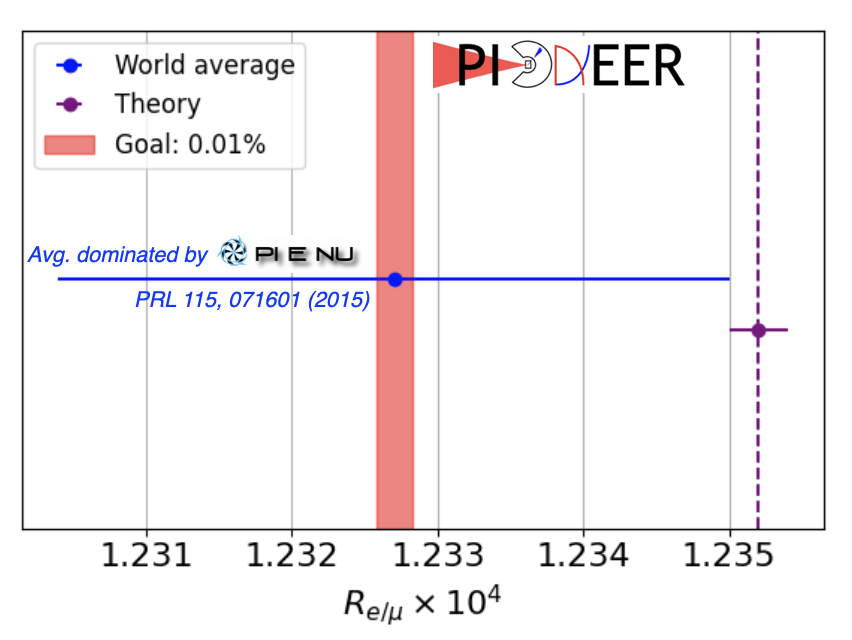} &
\includegraphics[width=0.48\linewidth, height=160pt]
{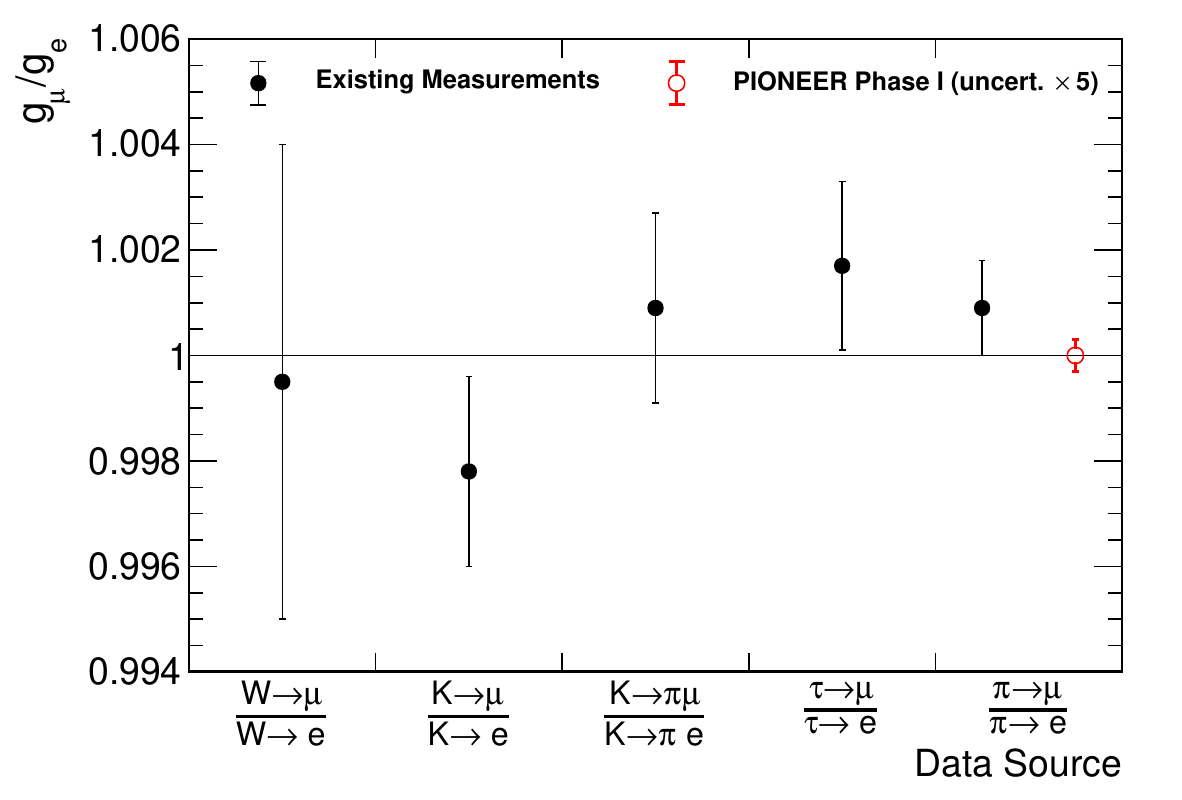} \\
(a) PIONEER Phase I goal & (b) Comparison of LFU tests  \\
\end{tabular}
\caption{(a) The figure summarizes the goal of the PIONEER experiment for the measurement of $R_{e/\mu}$.  The current world-average experimental value and its error bar are reported in blue. The theoretical prediction and its uncertainty are shown in purple. The red band represents the size of the projected PIONEER uncertainty.  (b) Comparison of probes of LFU tests between the first and second generation. Existing measurements from Kaons, Taus and Pions are taken from Ref.~\cite{annurev:vincenzo_doug_et_al_2022} and from Ref.~\cite{Aad2024ER-} for $W\to\mu/W\to e$. 
\label{fig:pioneer:remu}}
\end{center}
\end{figure}

Precise low-energy measurements of observables that can be very accurately calculated in the Standard Model (SM) offer highly sensitive tests of new physics (NP). The ratio $R_{e/\mu} = \Gamma(\pi^+\rightarrow e^+\nu(\gamma))/\Gamma(\pi^+\rightarrow \mu^+\nu(\gamma))$ for pion decays to positrons relative to muons is one such observables.
It provides a pristine test of Lepton Flavour Universality (LFU), a key principle of the SM. Tests of LFU can be conducted in others systems, such as tau decays, Kaons, B-hadrons or with on-shell W bosons but measurements of the charged pions are a particularly promising avenue. In the SM, $R_{e/\mu}$ is known with relative precision $1.2\times10^{-4}$~\cite{Cirigliano:2007xi}, making it one of the most precisely known observables involving quarks. 
However, the current experimental world average is about a factor 15 less precise  limiting the NP reach.
PIONEER is a new Rare Pion Decay Experiment~\cite{PIONEER:2022yag} approved with high priority at the Paul Scherrer Institute in Switzerland, where high-intensity pion beams can be delivered. In Phase~I, it  will bridge the gap of a factor 15 between theoretical and experimental precision for $R_{e/\mu}$, see Fig.~\ref{fig:pioneer:remu}. With measurements at the $0.01\,\%$ level in precision, NP up to the PeV scale~\cite{Bryman:2011zz} may be revealed. Such precision would  contribute to  stringent tests of LFU in a context where several intriguing hints of LFU violation (LFUV) have emerged.

PIONEER will probe modifications to weak interactions arising from various sources, such as possible new axial current or pseudoscalar currents, corresponding to a variety of ultraviolet models, including extended gauge and Higgs sectors 
(for more details see \cite{annurev:vincenzo_doug_et_al_2022} and references therein). 
Without committing to any explicit model,  the physics reach can be assessed 
using effective field theory considerations. 
Deviations from the Standard Model are captured by the ratio
\begin{equation}
    r = \frac{R_{e/\mu}}{R_{e/\mu}^{\rm SM}} = 1 + 2  \left( \epsilon_L^{ee} - \epsilon_{L}^{\mu \mu} \right)- 2 B_0 
    \left( \frac{\epsilon_P^{ee}}{m_e} - \frac{\epsilon_P^{\mu \mu}}{m_\mu} \right)
\end{equation}
where $B_0 (\mu_{\rm ren}) = M_\pi^2/(m_u(\mu_{\rm ren}) + m_d (\mu_{\rm ren}))$ and $\epsilon_L$, $\epsilon_P$
are low-energy charged-current effective couplings 
introduced in Ref.~\cite{Bhattacharya:2011qm}, corresponding to Standard-Model-like 
and pseudoscalar operators, respectively.  
Here $\mu_{\rm ren}$ denotes the renormalization scale in the low-energy theory 
describing the Standard Model augmented by the non-standard charged current interactions. 
Using the current experimental and theoretical input one finds  $r = 0.9980 (18)$, 
which implies  single-operator constraints at the level of $(\epsilon_L^{ee} - \epsilon_L^{\mu \mu})\sim 10^{-3}$ 
and $\epsilon_P^{ee} (\mu_{\rm ren} = 2~{\rm GeV})\sim 5 \times 10^{-7}$, the latter being greatly enhanced due to the lack of helicity suppression 
in the pseudoscalar amplitude.  Naively converting these sensitivities to new physics scales through $\epsilon_{L,P} = v^2/\Lambda_{L,P}^2$, where $v$ is the Higgs field vacuum expectation value, 
one obtains $\Lambda_L \sim 8$~TeV and $\Lambda_P \sim 350$~TeV. 
With the projected $O(10)$ improvement in precision obtained by PIONEER, the bounds on $\epsilon_{L,P}$ will 
improve by an order of magnitude and the effective scales $\Lambda_{L,P}$ will increase by $\sim \sqrt{10}$. 
This clearly shows that PIONEER will explore high-mass scales for new physics  well beyond the reach of direct HEP searches. 

The low-energy effective couplings $\epsilon_{L,P}$ can be expressed in terms of the Standard Model EFT Wilson Coefficients 
corresponding to dimension-six operators, normalized accordingly 
through ${\cal L}_{\rm SMEFT}^{(d=6)} =  \frac{1}{v^2}  \sum_i  C_i O_i^{(6)}$. 
Using the basis introduced in Ref.~\cite{Grzadkowski:2010es}, one finds (with the further specification that the up-quark Yukawa matrices  are diagonal~\cite{Dawid:2024wmp})
\begin{align}
    \epsilon_{L}^{ee} - \epsilon_L^{\mu \mu} &=  \left[ C_{Hl}^{(3)} \right]^{ee} -   \left[ C_{Hl}^{(3)} \right]^{\mu \mu} 
    - \frac{1}{V_{ud}} \left(  \left[ C_{lq}^{(3)}\right]^{ee11} -  \left[C_{lq}^{(3)}\right]^{\mu \mu 11}  
    \right) 
\\
\epsilon_P^{\alpha \alpha} &= \frac{1}{2 V_{ud}} \, \left[C_{ledq}^\dagger - C_{lequ}^{(1) \dagger} V \right]^{\alpha \alpha 11}~, 
\end{align}
where $V$ denotes the CKM matrix, the leptonic family indices are indicated by Greek letters,  and the remaining indices refer to  quark families. 
Modifications to the $W$ boson coupling to leptons of family $\alpha$ are encoded by $[C_{Hl}^{(3)}]^{\alpha \alpha}$. 
Focusing on this class of new physics interactions, one can compare the bounds on ratios of effective couplings 
$g_{\alpha} \equiv g (1 + [C_{Hl}^{(3)}]^{\alpha \alpha})$ from a variety of experiments, as shown in the right panel of 
Fig.~\ref{fig:pioneer:remu}.  This shows again the competitiveness of PIONEER.

\begin{figure}[h]
\begin{center}
\begin{tabular}{cc}
    \includegraphics[width=0.47\textwidth]{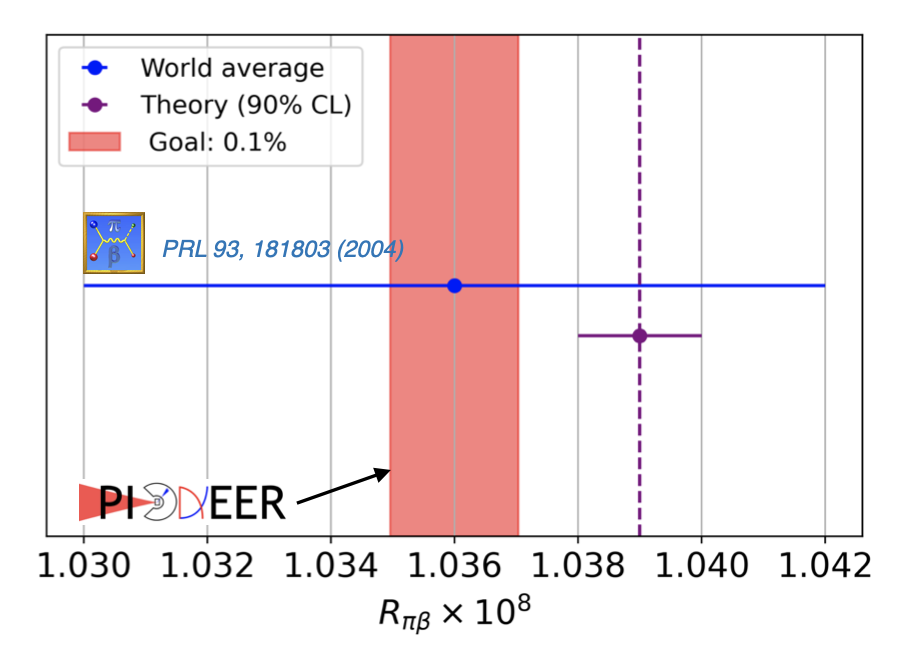} & 
    \includegraphics[width=0.47\textwidth]{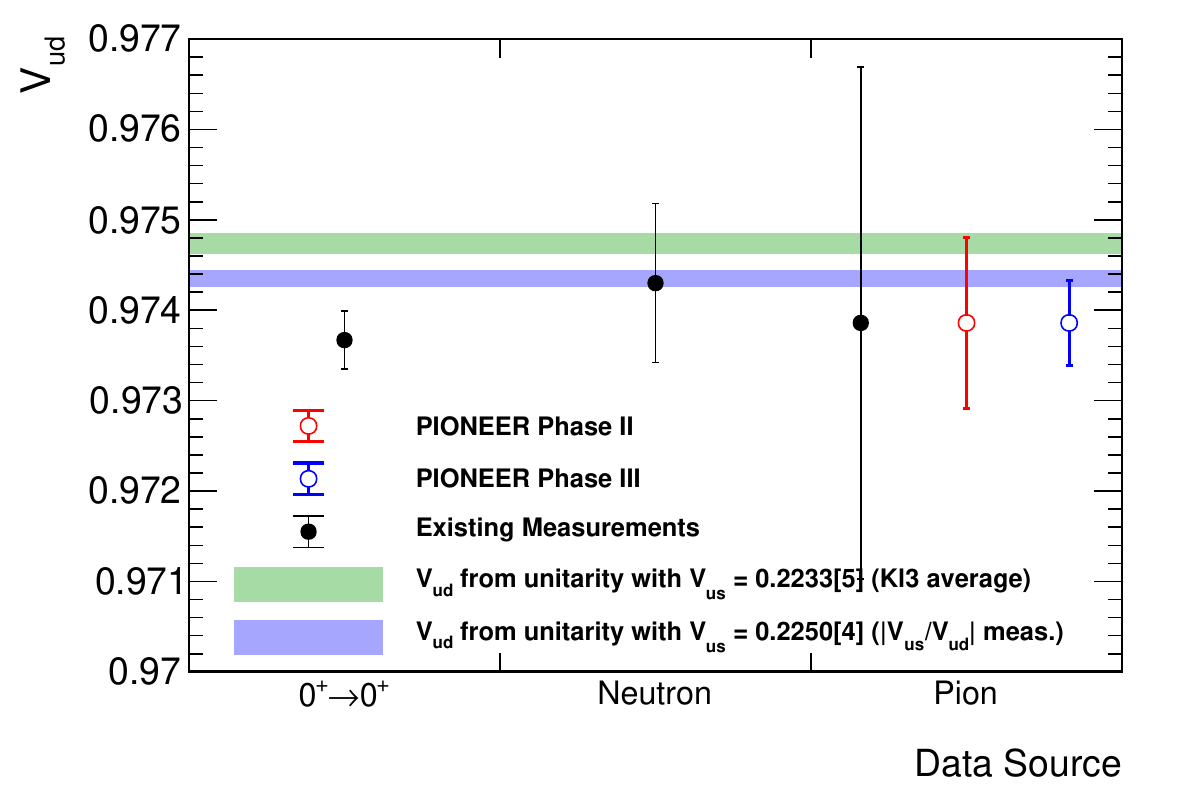}\\
    (a) PIONEER Phase II-III goal & (b) Comparison of $V_{ud}$ probes\\
\end{tabular}
\caption{(a) In Phase II-III, we plan to measure $R_{\pi\beta} = \Gamma(\pi^+\to\pi^0e^+\nu_e)/\Gamma(\pi^+\to \text{all})$. A factor of 3-, then 6-fold improvement in $R_{\pi \beta}$ will provide an independent and theoretically clean extraction of $V_{ud}$ in the CKM matrix. The current world-average experimental value and its error bar are reported in blue; the theoretical prediction is in purple; the projected PIONEER uncertainty is the red band. (b) Comparison of the current measurements of $V_{ud}$ with the estimate from CKM first row unitarity and the PIONEER Phase II and III projected precisions. PIONEER projections are preliminary and still under evaluation.
\label{fig:pioneer:pibeta}}
\end{center}
\end{figure}

In later Phases (II,III),  PIONEER  will  also study  pion beta decay $\pi^+\to \pi^0 e^+ \nu (\gamma)$ ultimately aiming at  an order of magnitude improvement in precision to determine $V_{ud}$ in a theoretically pristine manner and test CKM unitarity, for which there is presently a $ 3\sigma$ tension~\cite{ParticleDataGroup:2020ssz}, see Fig.~\ref{fig:pioneer:pibeta}.


Finally, the very large datasets collected by PIONEER will provide a platform to search for new feebly-coupled particles (such as heavy neutral leptons
or axion-like particles). For example, emission of a heavy neutral lepton (HNL) with mass in the O(10 -- 100)\,MeV range would produce an anomalous peak in the outgoing positron energy, and even if the HNL is too massive to be
emitted, the associated lepton mixing would generically cause the ratio
$R_{e/\mu}$ to differ from its SM value, thereby producing an
apparent violation of $\mu$-$e$ lepton flavor universality~\cite{PhysRevLett.130.241801,AGUILARAREVALO2019134980,PIENU:2017wbj,PIENU:2020las,PIENU:2020loi,PIENU:2021clt,PhysRevD.100.073011}.


PIONEER is an ambitious program that will span more than a decade of research activity. This document provides a brief overview of the detector concepts, simulations, estimated sensitivities and planning for realization. Ref.~\cite{PIONEER:2022yag} provides the most comprehensive description of the effort to date. The figures and values quoted in this section are meant to provide meaningful inputs to the Physics Preparatory Group, in particular for the working groups 1, 3, 4, 6, and 7. More details are available upon request. 

\section{The PIONEER experiment}

\begin{figure}[h!]
\centering
\includegraphics[width=\textwidth]{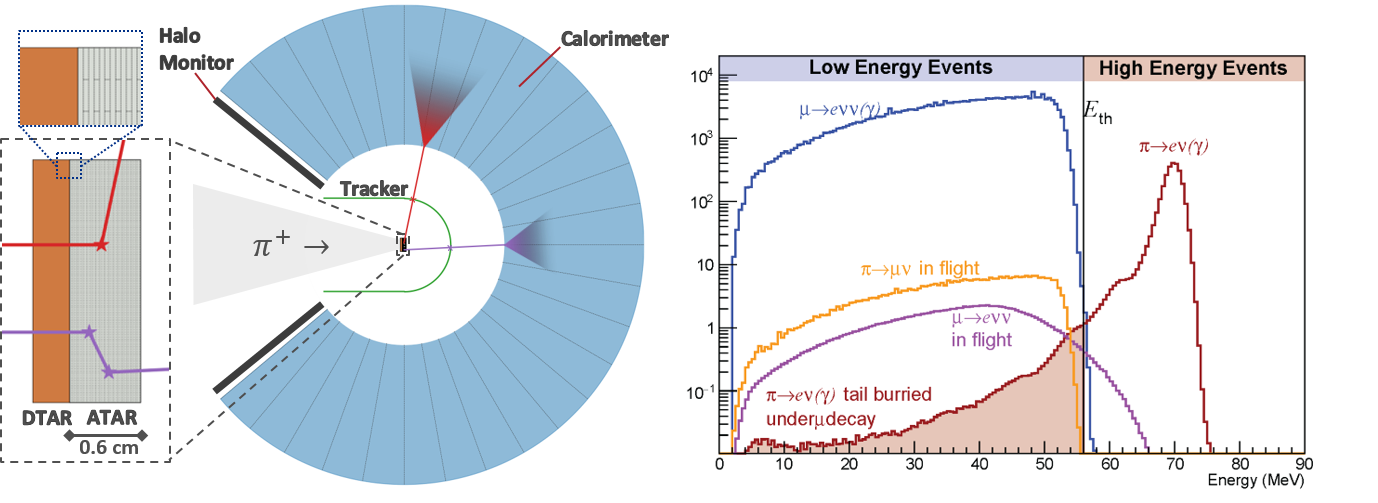}
\caption{Left: Layout of the PIONEER rare pion decay experiment.  The intense positive pion beam enters from the left and  is brought to rest in a highly segmented active target (ATAR).   Decay positron trajectories are measured from the ATAR to an outer electromagnetic calorimeter (CALO) through a tracker.  The CALO records the positron energy, time and location. Right: The positron energy spectra from  muon decays (blue) and from \pie~ decays (dark red). The (shaded) low-energy tail of the \pie shower lies below the Michel events in the Low-Energy Event bin, and also below decay-in-flight processes.
}
\label{fig:GenericDetector}
\end{figure}

The main challenge in developing a next generation experiment for a high precision measurement of rare pion  decays is  accurately assessing the performance of the chosen detector technology in suppressing sources of systematic uncertainties and handling increased rates. The PIONEER detector design concept, described in the next sections, is based on
the experience gathered with the PIENU \cite{PiENu:2015seu} and PEN/PiBeta \cite{Pocanic1,PEN:2018kgj, Pocanic:2014jka} experiments.
Generically, the detector will have the features sketched out in Fig.~\ref{fig:GenericDetector}.  An intense pion beam is brought to rest in an instrumented (active) target (ATAR) and an electromagnetic calorimeter (CALO) surrounds the stopping target. A cylindrical tracker surrounding the ATAR is used to link the locations of pions stopping in the target to showers in the calorimeter.
Features of the PIONEER approach will include improved time and energy resolutions, greatly increased calorimeter depth, high-speed detector and electronic response, large solid angle coverage, and complete event reconstruction. The proposed detector will include a 3$\pi$ sr, 19 radiation length ($X_0$) electromagnetic calorimeter, an advanced design segmented stopping target, and  beam and  positron trackers.

 \subsection*{Requirements for measuring $R_{e/\mu}$ }
 Phase I of PIONEER aims to measure $R_{e/\mu}$ with precision of 0.01\,\%, where the uncertainty budget is equally allocated to statistics and systematics; $2\times 10^8$ $\pi^+\to e^+ \nu$ events are required.

 At rest, the pion lifetime is 26\,ns and the muon lifetime is 2197\,ns.  The monoenergetic positron from $\pi \rightarrow e\nu$ has an energy of 69.3\,MeV.  Positrons from ordinary muon decay form the Michel spectrum from 0 to an endpoint of 52.3\,MeV.  In principle, the monoenergetic $e^+$ from \pie~ is well isolated above the Michel endpoint and can be easily identified using a high-resolution, hermetic calorimeter. To determine $R_{e/\mu}$, we measure the ratio of positrons emitted from \pie~ and \pimue~ decays for which many systematic effects, such as solid angle acceptance, cancel to first order. However, counting all \pie~ events with a precision of one part in $10^4$ requires determining the low-energy tail of the electromagnetic shower and radiative decays that hide under the Michel spectrum from the \pimue~ chain, which has four  orders  of magnitude higher rate. 
 
Figure~\ref{fig:GenericDetector} (right) illustrates the relationship between the two channels and their respective positron energy spectra.  Here, we have modeled the spectrum from both channels assuming a high resolution, $19\,X_0$ calorimeter with 1.7\% resolution at 70\,MeV. The unavoidable tail fraction below 53\,MeV must be determined accurately in order to obtain the branching ratio. That challenge was critical to previous generations of  experiments and was responsible for the leading systematic uncertainty in the PIENU experiment at TRIUMF. PIONEER will minimize the intrinsic tail fraction through the use of a high-resolution calorimeter having a uniform and large absorption fraction vs. polar angle, as illustrated in Fig.~\ref{fig:GenericDetector}.  Two technologies are presently being investigated.  One involves a continuous volume of Liquid Xenon, viewed by VUV photomultiplier and SiPM sensors on both the inside and outside surfaces.  This concept is based on the considerable experience of the MEG~II calorimeter, as documented in Ref.\cite{Mihara_2011}.  A second technology would employ 311 tapered LYSO crystals that form a closed array. Each crystal will be read out by an individual PMT.  A test with full-scale, tapered crystals is in progress. 


The branching ratio $R_{e/\mu}$ will be obtained by first separating the events into high- and low-energy regions using an energy cut value ($E_{cut}$) as was done by the PIENU experiment. The time spectra will be fit in each region with the $\pi^+ \rightarrow e^+ \nu$ and $\pi^+ \rightarrow \mu^+ \rightarrow e^+$ timing distribution shapes, along with  backgrounds originating from different sources including event pile-up effects,  pion decays in flight, and effects from old muon decays.

\begin{figure}[h!]
\centering
\includegraphics[width = \textwidth]{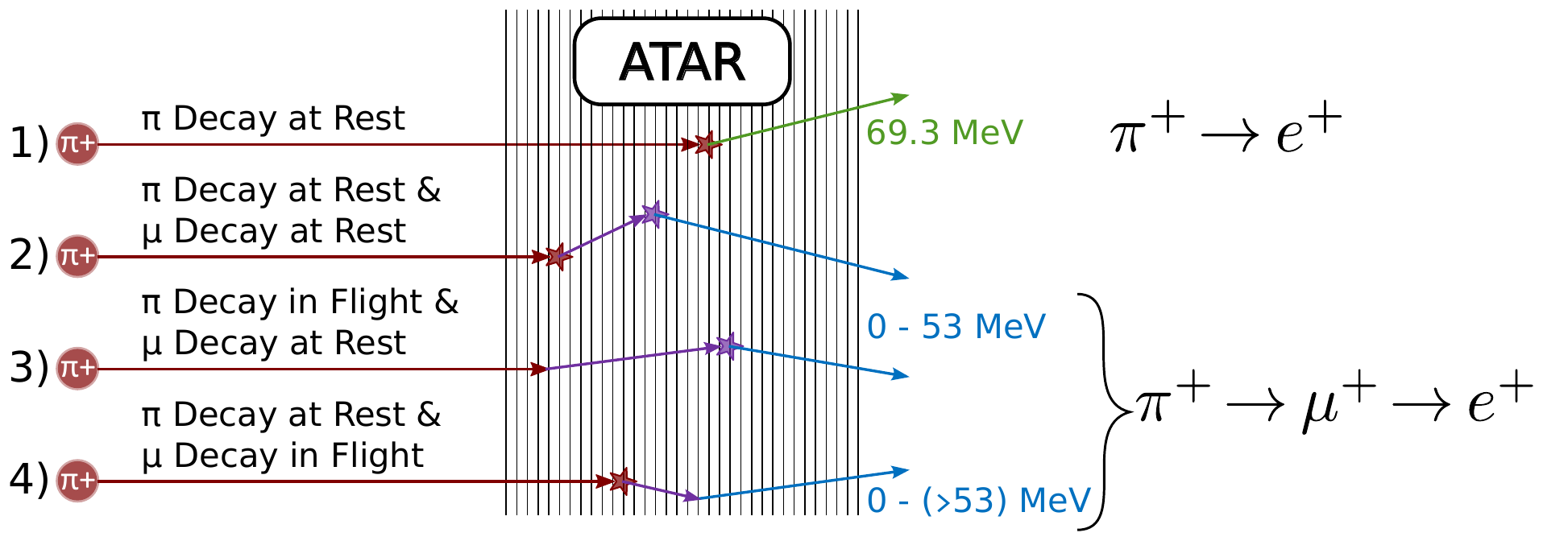}
\caption{Illustration of event types in the $\pi \rightarrow e$ and $\pi \rightarrow \mu \rightarrow e$ chains. The stars are indicative of the energy deposited at the Bragg peak for pions (red) or muons (purple) that stop in the segmented ATAR.  The main channels of interest include 1) the $\pi \rightarrow e$ ``signal channel'' decay that emits a monoenergetic 69.3\,MeV positron; 2) the dominant $\pi \rightarrow \mu$ decay, where the 4.1\,MeV muon travels up to 0.8\,mm and also stops in the ATAR before emitting a  positron.  Events 3) and 4) represent situations that can confuse the classification of events into categories 1) or 2). In 3) the pion decays within the ATAR  prior to stopping; the muon stop can then appear as a pion stop.  In 4), the pion stops, and the decay muon (very rarely, but importantly) decays in the short time prior to stopping. Because it is a decay in flight, the Lorentz boost can push the positron  energy beyond the 52.3\,MeV endpoint.  The ATAR is being designed to distinguish these event patterns.
}
\label{fig:EventTypes}
\end{figure}

Beyond the use of a calorimeter that has better energy containment, the stopping target must provide information to distinguish the event types shown in Figure~\ref{fig:EventTypes}.  Tracking, energy, and timing information will be used to distinguish $\pi \rightarrow e$ events from $\pi \rightarrow \mu \rightarrow e$ events, to identify pion and muon decays in flight, and to aid in identify pileup from long-lived muons that remain in the target from earlier pion stops.
Triggers will be developed to enhance the collection of various event types.

To distinguish event types, we will use an active target that can provide 4D tracking (at the level of 150\,$\mu$m in space  and $<$1\,ns in time) and energy measurements from the O(30)\,keV  signals for positrons to the 4000\,keV Bragg peaks of stopping pions and muons.  As discussed in Sec.~\ref{sec:ATAR}, our collaboration is focusing on the new low gain avalanche diode (LGAD) sensors as a centerpiece of the experiment.  Simulations using optimized LGAD parameters provide confidence that triggers can be constructed to isolate and measure all event types. 

The calorimeter tail fraction for \pie~ events  will be measured {\it in situ} by suppressing the \pimue~ decays using information provided by the active target. \pimue~ events can be identified and suppressed  by the presence of the 4.1\,MeV pulse from $\pi \to \mu \nu$ decay, use of a narrow time window, $\pi-\mu$ particle identification, and tracking information to identify pion decay-in-flight (\pdif), and muon decay-in-flight (\mdif) following pion decay-at-rest (\pdar).  Simulations indicate that $\mu-e$ backgrounds in the tail region can be suppressed to a level that will allow the uncertainty in the tail fraction to contribute $<0.01\%$ to the error in $R_{e/\mu}$.

The experiment will require a continuous wave pion beam that can be focused to a small spot size and  stopped within the ATAR dimensions.  Ideal characteristics include a relatively low momentum of $65\,$\,MeV /$c$ ($\pm2\%$) and a flux of 300\,kHz.  At this momentum, a separator is an effective method to reduce background from beamline muons and positrons.   We have already demonstrated that the $\pi$E5 beam can  provide the needed flux.  
Because of the high data rate, state-of-the-art triggering, fast digitizing electronics, and high bandwidth data acquisition systems are required.

\subsection*{Requirements for measuring pion beta decay }

In Phase II (III), pion beta decay $\pi^+\to\pi^0 e^+ \nu$ will be measured by observing the characteristic (nearly) back-to-back gammas from $\pi^0$ decay normalized to \pie~ decay as in Refs.\cite{Pocanic5,Frlez:2003vg,Pocanic:2003pf,Frlez:2003pe,Bychkov:2008ws}. In \nexp\ we also expect to observe the low-energy positron absorbed in the ATAR in coincidence with the gammas in the calorimeter.
The Phase II (III) pion beta decay experiment will require $7\times 10^5$ ($7\times 10^6$) events at an intrinsic branching ratio of $ 10^{-8}$.  This will require running at a significantly higher pion flux of $\geq 10$\,MHz.  The beam momentum and emittance may  be higher than for the \pie~ measurement to achieve the higher flux.  
The higher rate can be handled because of the  gamma ray coincidence identification and nearly fixed energy sum. The  $\pi$E5 beamline appears to have the necessary properties for this measurement.

\section{Imaging system}
\label{sec:ATAR}
The PIONEER imaging system consists of the \atar\ and the \tracker. 
\atar\ is a key enhancement over earlier experiments and crucial for achieving its precision goal in \remu. ATAR will define the pion stop region and suppress decay in flight and accidental positrons from muon decay to allow the experiment to run at its high beam rate. It is indispensable for suppressing the dominant \pimue\ chain by more than six orders of magnitude to directly measure the tail of the \pie\ positrons in the calorimeter response.
\atar\ is using an emerging detector technology known as low gain avalanche diodes (LGADs)~\cite{sadrozinski_4d_2018}, which are thin silicon detectors with a highly doped gain layer. Our research program aims to develop this technology into a unique 5-D tracking device featuring precision time information (\SI{0.2}{\ns} time and \SI{2}{\ns} pulse-pair resolution),  precision 3-D tracking at the \SI{100}{\micro\meter} scale and good energy resolution.  
The system will employ 48 individual LGAD strip sensors alternating in X and Y direction, each with an area of 20$\times$20\,mm$^2$ area and a thickness of 120 $\mu m$, stacked with minimal dead material in between planes. In total, 4800 channels will be read out with a large dynamic range (several hundred) and minimal cross-talk.
To reduce the dead material between ATAR and the calorimeter, the unamplified signal will be carried by roughly 15~cm-long low-mass flexible circuit boards from the sensor to outside the active region. The readout board using the FAST chip for amplification will be sitting at the edge of the calorimeter acceptance region. A fast digitizer system will fully digitize all the channels. Two digitizer options are taken into consideration: the HD-SOC from NALU Scientific and the SAMPIC from IJCLab.

An ATAR demonstrator is planned to be built by mid-2026 and tested by the end of 2026 at PSI. The updated demonstrator will then be tested at TRIUMF in 2027.
The demonstrator will be fabricated using TI-LGADs (trench-isolated) sensors from an ongoing common FBK production, FAST2 or FAST3 chips from INFN Torino, and a combination of the Sampic and HD-SOC digitizers. 
It will consist of 16 detector planes with a limited lateral size and around 500 total channels, roughly 10\,\% of the final ATAR.
The sensor and readout preparation and assembly will be exercised in 2025; a prototype composed of a few detector planes is foreseen to be tested at TRIUMF and CENPA by the summer of 2025.
Together with the ATAR demonstrator, a \tracker\ demonstrator is foreseen to be built in the same timescale.

The \tracker\ will determine the positron impact location on the \calo\ and identify pile-up events. It is designed for high hit detection efficiency (greater than 99\%) and nanosecond-level time resolution. The collaboration is exploring two potential technologies for the tracker: a low-mass Micro Resistive Groove ($\mu$RGroove) detector and a silicon pixel detector.
A tracker demonstrator is foreseen alongside the ATAR demonstrator.

The $\mu$RGroove option can have $\sigma_T<\SI{5}{\ns}$ and $\sigma_X<\SI{50}{\micro\meter}$, while maintaining a radiation length less than 1.0\% of the total radiation length.
It is planned to have a unique bullet-shaped geometry with the analog signal to be fed to the front-end cards using flexible cables. 
Tests with $\mu$RGroove prototype detectors and simulation and design studies will be continued to finalize the design, followed by the procurement and assembly of the demonstrator. 
The second option would be composed of depleted monolithic active-pixel sensors (DMAPS). DMAPS can be built very thin (thickness $<\SI{100}{\micro\meter}$), have a very good spatial resolution ($\sigma_X<\SI{50}{\micro\meter}$), and can be made fast ($\sigma_T<\SI{1}{\ns}$). 
A tracker built from such sensors would have a cylindrical shape, with or without a flat end-cap. 
DMAPSs are being developed by PSI \cite{Burkhalter:2024csp} with a focus on smaller experiments such as PIONEER. 
The first small prototype was delivered at the beginning of 2025, and a second larger prototype is scheduled for submission in early 2026. 

\section{Calorimeter}
The PIONEER Calorimeter (\calo) must accurately measure the positions, timing, and energies of particles emitted from the target, as well as any `pileup' resulting from beam particles decaying in flight or missing the target. For the $R_{e/\mu}$ measurement the \calo\ will measure positrons in the range from 0-$\sim$70~MeV. Essential characteristics of the \calo\ include a uniform response across different incoming positron angles, extensive coverage to enhance statistics and reduce acceptance corrections, excellent energy resolution, rapid timing, and short pulse duration.   
A critical aspect of the \pie~measurement is precisely determining the low-energy tail of the 70~MeV response function. 
Achieving PIONEER's precision target for $R_{e/\mu}$ requires minimizing the low-energy tail and accurately measuring the calorimeter's low-energy response to compare with Monte Carlo simulations. Simulations have shown that a calorimeter depth of ~19$X_0$ is sufficient to minimize contributions to the tail originating from leakages at the back of the calorimeter. Other contributions to the tail include photonuclear effects, albedo and lateral leakages.   

 Building a calorimeter that operates well in this low-energy regime is challenging. The considered options include a monolithic (unsegmented) sphere of liquid xenon (LXe). The MEG Collaboration~\cite{MEGII:2023fog} has already demonstrated excellent in-its-class energy and position resolution for a 53\,MeV gamma that would emerge from the cLFV decay $\mu \to e\gamma$ using a LXe calorimeter and those experts are playing a central role in designing the PIONEER LXe concept. An alternative approach is based on tapered LYSO crystal segments that together form a truncated polyhedron geometry in a more conventional structure. There are different advantages to both approaches.
\subsection*{LXe approach} 
The LXe option is inspired by the PSI MEG~II experiment \cite{MEGII:2023fog}. This approach is novel - previous pion decay experiments have used single or arrays of crystal detectors. A monolithic detector allows the containment of the electromagnetic showers in a single homogeneous volume leading to high energy resolution and minimization of dead material effects. 
LXe has excellent properties: high density,  $\textrm{X}_0 = 2.87$\,cm; $\textrm{R}_M = 5.22$\,cm; light yield $\sim 65,000 \gamma$/MeV,  $\tau_{decay} \sim 40$\,ns, \SI{65}{\pico\second} time resolution.
These properties make LXe well suited to precisely measure particle energies in a high rate environment. The current geometry consists of a partial sphere containing 4.5 tons of liquid xenon. The scintillation light is collected by photosensors uniformly distributed across the surface of the calorimeter: chip-on-film (CoF) silicon photomultipliers (SiPMs) on the inner surface of the detector and photomultiplier tubes (PMTs) on the outer surface. This configuration ensures very good resolution for incoming decay positrons. The CoF SiPMs, assembled on thin hemispherical windows, are currently at the research and development stage. A large LXe prototype ($\sim 650$~kg of LXe) is in preparation to test the performance of such assembly and benchmark optical simulations. Tests with a positron beam at PSI are envisioned in 2026.

\subsection*{LYSO approach} 
A LYSO crystal design results in a relatively simple event reconstruction and reduced pileup because only a few crystals are illuminated when a positron strikes the calorimeter; together they create compact clusters with summed energy and averaged time.  Important LYSO characteristics include:  $X_0 = 1.14$\,cm; $R_M = 2.07$\,cm; light yield $\sim 32,000 \gamma$/MeV, $\lambda_{peak} = 420$\,nm, $\tau_{decay} = 40$\,ns. LYSO is radiation-hard and not hygroscopic.  These characteristics are attractive, but the performance for large crystals had not yet been demonstrated at the level required for PIONEER.  Over the last 2 years, we have been working with the Shanghai Institute of Ceramics to develop next-generation LYSO crystals, optimized for particle physics. In work recently published~\cite{Beesley:2024mts} we document our findings using an array of ten rectangular-shaped crystals.  We measured the energy resolution of 1.5\% for 70\,MeV positrons and 110\,ps timing for events above 30\,MeV. These performance characteristics exceed the requirements for PIONEER.  The next stage of R\&D is devoted to the testing of full-scale, tapered crystals.  Three have been produced to date; each demonstrate excellent resolution using bench sources and an intrinsic longitudinal response uniformity of better than a few percent.   A beam test at PSI with a small tapered array is scheduled for August 2025.

\section {Collaboration composition}
The PIONEER collaboration consists of experts in high-precision experiments including rare pion and kaon decays, a search for charged lepton violation, and the measurement of the anomalous magnetic moment of the muon. In addition, we have collaborators from the high-energy collider and neutrino communities.  Critically, we emphasize that the collaboration is young and still developing and warmly welcomes new members. European institutes are ideally positioned to contribute to the project, as it will be located at the PSI laboratory in Switzerland.

\section{Timeline}

\begin{figure}[h!]
\begin{center}
\includegraphics[clip, trim=2.25cm 16.5cm 2.25cm 1.8cm, width=1.00\textwidth]{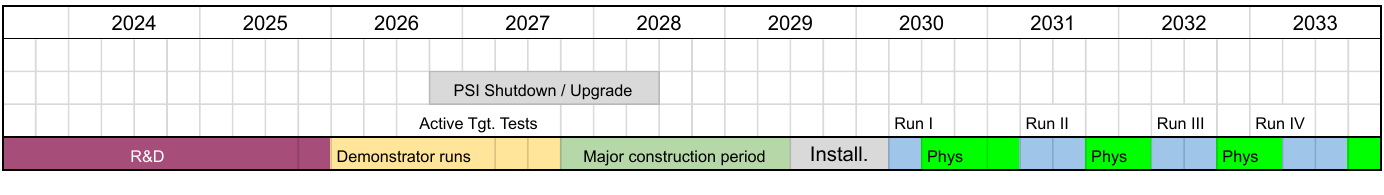}
\caption{Timeline of the PIONEER project.\label{fig:timeline}}
\end{center}
\end{figure}

Fig~\ref{fig:timeline} illustrates an optimistic schedule for \nexp\, under the assumption that funding decisions are positive and proceed expeditiously.  Current efforts are focused on critical demonstrations of subsystems (e.g. ATAR, Calo, Traker, DAQ), which will be tested in the upcoming years. Calorimeter test beam time and piE5 beam development tuning is planned for 2025 and 2026.  Demonstrator runs are expected at PSI in late 2026 and at TRIUMF in 2027. Positive outcomes will lead to full-scale construction and a measurement program beginning in 2030.  

\section{Construction and operational costs}

Several costing exercises were performed for \nexp\ and are available upon request. The overall cost of the experiment is estimated to be $10-15$ M\euro{}. The calorimeter is the most expensive part of the system. For the LYSO option, manufacturing the crystals dominates the costs while for the LXe, procurement of LXe is notoriously volatile and the instrumentation will also be a significant fraction of the cost. The active target is estimated around $1-2$M\euro{} with the digitisation system being the most expensive component. 

The Paul Scherrer Institute has a strong track record of hosting long-term, precision experiments in pion and muon physics. 
They provide the beams without cost, and the support of beam lines, experimental counting rooms, work areas, installation personnel, and technical support as needed to ensure success of approved experiments.
The PSI Program Advisory Committee, in turn, closely monitors the progress of all experiments throughout the life cycle from design to completion and results.

\printbibliography
\end{document}